\documentclass{article}
\usepackage{ijcai19}

\usepackage{times}
\usepackage{soul}
\usepackage{url}
\usepackage[hidelinks]{hyperref}
\usepackage[utf8]{inputenc}
\usepackage[small]{caption}
\usepackage{graphicx}
\usepackage{amsmath}
\usepackage{booktabs}
\title{Science Communications for Explainable Artificial Intelligence}

\author{
Simon Hudson$^1$\footnote{Contact Author}\and
Matija Franklin$^2$\\
\affiliations
$^1$BottoDAO\\
$^2$University College London\\
\emails
\ hudsonsims@gmail.com
}

\begin{document}

\maketitle

\begin{abstract}
 Artificial Intelligence (AI) has a communication problem. XAI methods have been used to make AI more understandable and helped resolve some of the transparency issues that inhibit AI's broader usability. However, user evaluation studies reveal that the often numerical explanations provided by XAI methods have not always been effective for many types of users of AI systems. This article aims to adapt the major communications models from Science Communications into a framework for practitioners to understand, influence, and integrate the context of audiences both for their communications supporting AI literacy in the public and in designing XAI systems that are more adaptive to different users.
\end{abstract}

\section{Intro}

Artificial Intelligence (AI) is increasingly implemented in software products employed by users with a wide variety of backgrounds and competencies. As a new technology, its effective use depends on a great amount of effort to educate users in a way that enables them to adapt the tool to their needs. Making use of AI is made more difficult due to a lack of transparency \cite{lawrenceband}, and hoping black boxes will align with our human goals is not a good alternative \cite{gabriel2020artificial}. To ensure that humans have effective oversight over AI agents, Explainable Artificial Intelligence (XAI) techniques have been developed to help users make sense of an AI's decisions and predictions. 

However, XAI models provide primarily numerical explanations favored by those with numerical backgrounds \cite{ehsan2021explainable}. For example, SHapley Additive exPlanations (SHAP) - a method for explaining individual AI model decisions by computing the contribution of each variable to the decision - can lead to information overload for lay users \cite{ferguson2022explanations}. The same research found that simply written messages that provide a list of factors that were considered by the AI model increased trust and understanding. There is a dilemma where the SHAP model objectively provided more information, but a written message provided as an alternative was perceived as more understandable and trustworthy. 

It appears that AI has a communication problem. There is scarce evidence for what explanation should be given to a certain individual for a specific task \cite{franklin2022influence}, and therefore it is not clear how to optimize the distribution of XAI models \cite{franklin2022human}. We argue that XAI, and the AI field more broadly, can benefit from the Science Communications (SciComms) field to better integrate missing user context. SciComms is the practice of communicating scientific information to the public in an accessible and engaging way. It involves researching, writing, and editing scientific content, creating visuals, and using various digital and traditional communication tools. The strong public discourse around AI's impacts sparked by Chat-GPT calls for greater use of practices developed in the SciComms field to help build AI literacy; generative text models also present an opportunity to integrate the SciComms approaches into AI interfaces directly to improve their adaptability, and usefulness, to different audiences. 

\section{AI's Communication Challenge}

Enabling users to effectively make use of AI has largely been treated as an AI literacy problem: users having sufficient understanding of the technology's functioning in order to "critically evaluate AI technologies; communicate and collaborate effectively with AI; and use AI as a tool online, at home, and in the workplace \cite{long2020ai}." Treatments have followed a communication model in SciComms referred to as the "information-deficit model". This model assumes what users are missing are facts about a given field of science or technology, and addresses that gap through top-down dissemination of information. The information deficit model is typically a top-down, one-way communication approach that simply relies on simply disseminating the knowledge of experts. Even if the ideas are simplified, they tend to ignore the individual contexts and local knowledge of users and instead favor a one-size-fits-all approach. These efforts tend to fail in impacting the overall literacy of a field due to not adapting to the audience's various perspectives or being made relevant in a way they can find useful in their daily lives \cite{miller2001public}.

SciComms encompasses the "skills, media, activities, and dialogue to produce" robust engagement with science and technology in the public \cite{burns2003science}. For a technology like AI, SciComms is a critical practice in preparing the public for its effective use. The SciComms field has been evolving out of the “information deficit model" that pins mistrust in science on science illiteracy \cite{nisbet2009s}. Meta-analysis of surveys on the public understanding of science taken around the world since 1989 has shown that trust in science varies even as average science literacy stays the same \cite{allum2008science}. Other contexts \cite{nisbet2007understanding}, in particular, historical experiences with science institutions \cite{wynne1992misunderstood}, are much stronger predictors. Understanding the context of past influential experiences, as well as the context of the present narratives and issues around technology, helps determine the frames most effective at supporting productive discussion in the public \cite{nisbet2009s}. These contexts are a moving target as AI makes more of an impact on mainstream culture due to the release of widely publicized products like Chat-GPT. 

Other models in SciComms have been growing in favor, including the \textit{contextual model}, \textit{lay expertise model}, and the \textit{participatory model} \cite{brossard2009critical}. The contextual model emphasizes adapting communications to different audiences' diverse contexts that can include a wide variety of factors: such as attitudes, beliefs, the task being executed, the domain it is performed in, and the current circumstance of a user. The lay expertise model gives additional credence to local knowledge, such as cultural or domain expertise that the science and technology experts do not have access to; it balances the authority of the science and technology expert with that of the end-user and knowledge of their own needs. The participatory model emphasizes the key role of holders of local knowledge in forming goals for technology design and even the policy regulating it. The deficit model has not been entirely discounted, rather the SciComms field has pursued balancing it with the strengths of these other models. 

Resolving AI's communication challenge thus should include the goal of better understanding different stakeholders by exploring the contexts most relevant to them so that they can give input they believe is valuable. Further, AI outputs themselves are a kind of scientific communication of the data science the models employ to make a prediction, and XAI would also benefit from applying these methods directly in an AI interface's presentation of outputs. Recent advancements in generative AI, as they apply to XAI, provide a valuable opportunity to integrate these models for SciComms into the design of \textit{adaptive explainable AI} to address the moving target of individual user contexts. 

Taking this view, AI's communication challenge can be divided into two categories: creator and user. The user challenge is in preparing users’ expectations of these interfaces via AI literacy efforts - communicating the dynamics and implications of algorithms so that people can raise their ability to think critically about the data generated and outputs provided by algorithms. The first creator challenge is in designing systems that are well adapted for their intended audience, or, if for general use, can be adaptable to a wide variety of users. The second creator challenge is in developing systems that are capable of learning and adapting to an individual user’s context by taking into consideration an individual's background, the task they are performing, the domain, the ordering of tasks, stated preferences etc. For example, explaining an AI's decision to a junior educator may be more detailed and authoritative versus a senior one who may receive only more nuanced explanations. This is of course complicated by other contexts, like their particular AI literacy, experience with the machine, or cultural factors. 

We propose a framework for addressing AI's communication challenge through three stages of work: Understanding Context, Influencing Context, and Integrating Context.

\section {Understanding Context}

Understanding Context involves gathering and interpreting knowledge of the target audience, their needs, values, and interests. It means understanding the user's background, their level of familiarity with AI concepts, their cognitive style, and their information needs. A nationally representative survey of 2,000 American adults in 2019 found that most people use AI, but claim that they don't and won't in the future \cite{zhang2019artificial}. Presenting AI as a consumer product (e.g., Netflix recommender system) versus as an intelligent agent influences people's receptiveness to it. On the other hand, giving AI agentic properties can obscure the human role in AI systems \cite{epstein2020gets}. It is therefore important to research how different contextual factors can set reasonable expectations of the technology and the public's own sense of agency in influencing its development \cite{bao2022whose}. 

Methods used in SciComms for understanding context involve audience research (i.e., understanding knowledge, attitudes, and interests), psychographics (i.e., understanding psychological attributes), demographics, narrative (i.e., understanding context through users' stories), and user testing (i.e., empirical research on what types of explanations work best for different types of users) \cite{van2018communication}. In addition, a growing set of computational approaches for measuring audience reception of communications will be helpful here \cite{august2020writing}. 

An interesting case study on how context can be understood, and how it influences the perception of emerging technologies comes from research on self-driving vehicles (SDVs). One notable factor is the lack of information provided by SDVs \cite{buscher2009intelligent}. A factor that may also reduce trust in SDVs is the widespread data collection that is necessary for a connected network of SDVs to work efficiently \cite{buscher2009intelligent}, however recent evidence suggests that many appear willing to share data if it improves travel experience \cite{wockatz2015traveller}. Further, individuals prefer SDVs whose algorithms are set in a way so that they protect their passengers at all costs, yet they would prefer if others bought vehicles that are programmed to sacrifice passengers for the benefit of the majority (e.g., saving a higher number of pedestrians; \cite{bonnefon2016social}). More broadly, people seem to be concerned about hacking, misuse, legal issues, and safety \cite{kyriakidis2015public}. Although it is evident that a car's ethical behaviour is outside the scope of SciComms, there is a clear opportunity to provide more information, and to be more transparent about what type of data collection is necessary for a safer and improved travel experience. Further, it can help inform which context to prioritize when determining a communication strategy.
 
SDVs also give insights into how sociodemographics are relevant to public attitudes. Studies have identified that younger travelers \cite{ruggeri2018new} and those living in large cities \cite{cohen2017social} are more willing to adopt SDVs. Further, people in China and India had more positive opinions about SDVs than people in the US, UK, and Australia \cite{schoettle2014public}. Identifying the sociodemographic divides in attitude is indicative of the forthcoming potential divides in the challenges SciComms about AI will face.

Finally, \cite{ruggeri2018new} found that attitudes towards SDVs reflect the theoretical curve assumed in Diffusion of Innovations Theory \cite{rogers2014diffusion}. SDVs, like other emerging technologies, will first be used by early adopters, successively followed by second-wave adopters, mainstream users, late adopters, and avoiders. The patterns exist within and between every age breakdown, however, the older participants were more likely to resist SDVs. Younger generations, considered digital natives, are more trustworthy of intelligent systems and algorithmic decision-making; thus, as a group, they skew towards early adoption. SciComms initiatives, thus, will have different challenges for each group of adopters. The movement of users between these categories as the technology spreads will also add value to strategies that can be adaptive to shifting contexts.

We are considering a broad definition of context, involving any variety of factors that can influence how information is received, interpreted, and integrated or discarded. Put simply, it is considering the audience's diverse needs in order to reach a communication goal and goes beyond a communicator only considering the information that they want to be known by the audience. To that end, considering different frameworks for understanding relevant context can be useful in identifying the priority factors to consider in the basic design of an XAI system. One approach to understanding the context in relation to XAI comes from \cite{franklin2022human} who argue that people's reaction to explanations can be evaluated by measuring their \textit{mental models, probability estimates, trust, knowledge, and performance}. Here, \textit{mental models} are representations of how people understand systems. \textit{Probability estimates} are the probability of future events occurring. \textit{Trust} is a multidimensional concept predictive of whether people choose to use AI. \textit{Knowledge} can be procedural - relating to changes in abilities - or semantic - relating to changes in factual knowledge. Finally, people's \textit{performance} on the task an AI is assisting them with will change due to the provided XAI explanations. The approach can identify what factors are the main barriers to communication. Some users may not trust the AI in the first place, while others may not have the procedural knowledge to use the tool.

\section{Influencing Context}

 Influencing Context involves shaping communication about AI (the user challenge) as well as the baseline design of interface features to address the specific needs and concerns of a target group (the first creator challenge). This means not only delivering explanations that the audience can understand but also considering what issues and consequences might be most relevant to them. 

"Framing" is a key practice used in SciComms to influence public reception of scientific and technological topics. A frame is a storyline selected to help simply convey complex topics. Different frames prioritize different issues and consequences, guiding attention to what the communicator wants an audience to focus on out of the many topics within a complex set of information \cite{nisbet2009s}. However, it is important SciComms practitioners make sure to avoid \textit{emphasis framing} - the phenomenon in which the bias of presented information also biases judgments of that information \cite{druckman2001implications}. Such framing can be used to influence a person's perceptions of an issue, either positively or negatively. For example, trust in the authenticity of AI work can be increased by emphasizing human involvement in their creation or training processes \cite{jago2019algorithms}. The goal of framing should be to draw their attention to the most relevant aspects of their lives so that they can productively engage with the ideas, without biasing the judgments they go on to make. 

Picking which aspects of technology to focus on can influence whether there is any effect on public perception and understanding at all. Studies have found that for emerging technologies, such as carbon nanotubes and genetically modified foods, factual information about the technologies has a lower impact on people's reception of such technologies than background factors such as people's values \cite{druckman2011framing}. The same study found that once people form their opinions, they tend to view factual information about a technology's features through \textit{motivated reasoning} - where they use this information to reaffirm their opinions. Even if someone wanted to emphasize features that they themselves see as positive, a receiver with a negative opinion may not be influenced at all. Framing that first addresses the storylines behind the beliefs may be a more effective approach.

Another approach to influencing context is message design. Message design involves designing the message to be engaging, understandable, and memorable \cite{livingstone2019audiences}. This can involve visuals, analogies, or narratives to make explanations more engaging, or repetition and reinforcement to make messages more memorable \cite{nisbet2009s} Audience research approaches, such as focus groups, can help inform and test narratives needed to reach specific groups \cite{krause2019trends}. Gaining such an understanding will also inform the development of AI technologies as they will emphasize user needs \cite{ruzi2021testing}. For example, when individuals are given a certain amount of control over intelligent systems, they are more likely to use them \cite{dietvorst2018overcoming}. 

While the SciComms field has discounted the deficit model, it has not completely thrown it out. It is still important to provide groups with details about science and technology. Public stakeholders respond strongly to being provided information on the technology’s functioning as well as broader implications in society \cite{besley2008interpersonal}. Integrating with the other SciComms models is a matter of incorporating feedback and adjusting the message. This involves monitoring the audience's reaction to the communication and adjusting it as needed. For example, if the audience is not understanding the explanation, understanding whether it comes from a knowledge gap or a communications gap is important. Further, SciComms can utilize issue prioritization -  identifying the most important issues and consequences for each group and focusing the communication on these issues.

SciComms can also influence AI literacy through a participatory approach. Participatory approaches include dialogues and even giving stakeholders access to decision-making discussions on design and research priorities. This not only helps ensure a deep integration of context in the research and development of the technology, but the process also better equips stakeholder groups to engage with and critique the efforts of the field \cite{boyd2012critical} and better present perspectives that can actually be taken up in development \cite{franklin2022causal}. It is important that these “bottom-up” strategies happen early and often, not only so that the public’s views, local knowledge, and lay expertise can be incorporated, but so that the public feels a part of the process, and consensus and reasonable expectations can be built before deployment \cite{etzioni2017incorporating}. This can also help supersede unhelpful or bad-faith framing. This practice has begun to a degree \cite{royal2020}, but we would emphasize that collective dialogues are as much a part of influencing context as they are about understanding context \cite{long2020ai}. 

Indeed, recent approaches have looked at how explanations provided by XAI can either nudge behaviour directly or boost people's capability which in turn promotes better AI use \cite{franklin2022influence}. From this perspective, local feature importance explanations and concept-based explications are harmonious to disclosure nudges that give context-dependant information, and global feature importance explanations and counterfactual explanations are congruous to boost capability. A further inquiry would allow for a more optimal distribution and selection of XAI methods and corresponding SciComms techniques. Translational exploration utilizing already current understandings from AI, Behavioural Science, and Human-Computer Interaction would be advantageous.

SciComms can help in the development of better human-in-the-loop (HITL) architectures, where ML systems communicate to humans via XAI explanations, and humans give feedback to update the model's performance \cite{bai2022training}. HITL ML architectures are systems in which humans and machines collaborate to create solutions \cite{zanzotto2019human}, using feedback loops where humans and machines interact and learn from each other \cite{mosqueira2022human}. This type of architecture is particularly useful in cases where the information and data available are not enough to provide an accurate solution. By combining human insight with computer algorithms and datasets, a more accurate and comprehensive solution can be created.

\section{Integrating Context}

 Integrating Context focuses on creating an interactive and adaptive communication interface that allows for a two-way dialogue between AI and people. The goal here is to establish a feedback loop where the AI learns from user interactions and modifies its explanations accordingly employing SciComms models, and where the user learns from the AI and adjusts their expectations and understanding of AI behaviour. Such communication is necessary to maintain human agency, and integrating techniques for context discovery and adaption in AI interfaces is an important topic of research. In other words, this is about extending the first two stages of Understanding and Influencing Context into an adaptive interface. To this end, computational linguistics is beginning to use science communications principles in conversational interface design that considers sensitive vocabulary when explaining a scientific concept so as to not run into a bias of a user, or to create a harmful one \cite{kocielnik2021designing}. Generative AI offers a large opportunity to do this adaption for the wide variety of context factors needed to consider in a globally distributed technology \cite{august2020writing}.

The SciComms field has made strides to update itself to the modern needs and media paradigms and incorporated more computational methods that would better enable application to the fast-moving and widely distributed field of AI \cite{freiling2021science}.

SciComms calls two-way dialogue \textit{dialogic communication}. This means the AI would not just provide explanations, but also solicit feedback, ask clarifying questions, and adapt its responses based on user inputs. A case study for the benefits of two-way communication comes from preferences inference and elicitation. \cite{russell2019human}  has argued that one way forward is to align AI to human preferences and that they can be best inferred from human behaviour. This perspective, however, ignores the fact that preference and behaviour are often influenced by the same factors \cite{franklin2022recognising}, and that preference and behaviour have a bidirectional causal relationship \cite{ashton2022problem}. As a result, we might want to design systems that attempt to elicit preferences or meta-preferences more directly by prompting users for information \cite{ashton2022solutions}. 

Another SciComms approach that may be relevant is \textit{adaptive communication} - AI dynamically adjusts its explanations based on the user's feedback. For example, if the user indicates they didn't understand an explanation, the AI could provide a simpler explanation or use a different analogy. The approach can be paired with an iterative design for XAI models - continuously refining the AI-user interface based on user feedback and testing results. This would result in a cyclical process of design, testing, feedback, and redesign, aimed at continuously improving the communication between the AI and the user.

One risk here is that a battery of questions probing for user info may result in disengagement of the user. Providing explanatory features that prompt more questioning from the user may be beneficial here. In a generative interface, for example, showing the hidden prompts that guide its answer, showing sources, or indicating the effect of increasing the temperature on the likelihood of inaccurate hallucinations can help the user to discern how an output was reached and ask follow-up questions that imply their context. Coupled with a simple suggestion to question the output and provide more context could prompt the user to offer more useful variables for the AI agent to adapt to. Larger token sizes of AI models would also increase the memory of these systems to learn and adapt to their users over time and tune them to their needs. 

Other risks included an AI model getting stuck in a particular tuning to a user and struggling to update to a user making significant changes in attitude or experiencing a large change in context. The user may also become overconfident in the AI system and stop questioning the outputs. For both user and machine it is ideal to maintain a healthy level of communicated uncertainty, and defining parameters for measuring the success of these systems will be a critical area of research. 

\section{Conclusion}
AI has a communications challenge. To create human-centric AI, it is important that XAI is able to adapt to different users. This has often been confused with trying to increase users' trust in AI without understanding why they might not trust it in the first place. SciComms is a field with rich insights into how to understand users so as to improve public engagement and expectations of AI systems, and its approaches could be employed to help AI systems better adapt to their particular users. 

Such integration can result in more adaptive XAI systems that mimic the way a skillful human communicator would adapt their explanations to different audiences in different contexts. Building systems that have that ability of explanation adaptation is now possible, where an explanation can be completely changed while retaining the inherent meaning. We argue that XAI needs to move towards \textit{adaptive explainable artificial intelligence } - a recommender system that predicts the preferred explanations of a specific user in a particular context. 

There is a wide area of research to pursue, and we propose a study of research gaps to identify useful areas of overlap between SciComms and XAI. To build adaptive XAI, data would be needed on how certain people react to different explanations while performing distinct tasks in particular contexts. There have been recent proposals for XAI evaluation paradigms \cite{franklin2022human}. Using the data collected from such paradigms in an adaptive system would involve building a simple, interpretable model capable of recommending the right XAI method to the right person for a given task. That data would contain well-understood variables, so a simple model can also produce salient explanations of its own process by default — a kind of ‘meta-interpretability' that stops new black boxes from hindering human-AI interaction.

\bibliographystyle{named}
\bibliography{ijcai19}

\end{document}